\documentstyle[epsf,12pt]{article}
\begin{document}
\begin{titlepage}
\begin{center}
\vspace*{3cm}

\begin{title}
\bold
{\Huge
Bose-Einstein effect in $Z^0$ decay and the weight method
 }
\end{title}

\vspace{2cm}

\begin{author}
\Large
K. FIA{\L}KOWSKI\footnote{e-mail address: uffialko@thrisc.if.uj.edu.pl},
R. WIT\footnote{e-mail address: wit@thrisc.if.uj.edu.pl}

\end{author}

\vspace{1cm}

{\sl M. Smoluchowski Institute of Physics\\ Jagellonian University \\

30-059 Krak{\'o}w, ul.Reymonta 4, Poland}

\vspace{3cm}

\begin{abstract}

We discuss the Bose-Einstein interference effect in multiparticle production. After a 
short review of various methods of implementation of this effect into 
Monte Carlo generators the weight method is presented in more detail and used to 
analyze the data for hadronic $Z^0$ decays. In particular, we consider the possibility 
of deducing the two-particle weight factor from the experimental data.

\end{abstract}

\end{center}



\vspace{1cm}

\noindent

August, 1999 \\

\end{titlepage}
\def\w{\vrule height 15pt depth 2pt width 0pt}
\section{Introduction}
   In the last years many papers have been devoted to experimental and
theoretical studies of Bose-Einstein interference effects~\cite{hbt}  in multiparticle
production. It has been argued that these studies may allow for the
reconstruction of space-time development of the interactions. In particular,
different possibilities of implementing interference effects into Monte
Carlo generators used for high energy hadroproduction processes were
discussed. In this note  we add some points to this discussion.
\par
   We present in detail some aspects of the weight method of implementing 
BE effect in the MC generators. In particular, we try to establish to 
what extent one may reconstruct the two-particle weight function 
(related to the Wigner function) from the data on BE effect. For this 
purpose we use the data on multiparticle production with highest statistics 
available - the hadronic decays of $Z^0$'s produced in $e^+e^-$ collisions.

\par
A short summary of various methods  of implementing
 interference effects into Monte Carlo generators is presented in the next
section. In the third section we discuss the data from different LEP
experiments on two-particle correlations from $Z^0$ hadronic decays which
were used for analyzing the Bose-Einstein effect.  Section 4 is devoted to
the analysis of data in terms of the weight model.  In particular, various
choices of two-particle weight factors used in this method are compared.
  Last section contains some conclusions and outlook for further
investigations.
\section{Implementation methods}
The standard discussion of the BE
effect in multiparticle production~\cite{bgj} starts from the classical space-time
source emitting identical
bosons with known momenta.  Thus the most natural procedure is to treat the
original Monte Carlo generator as the model for the source and to symmetrize the
final state wave function~\cite{sul}.  This may be done in a more proper way
using the formalism of Wigner functions~\cite{zha}.  In any case, however, the
Monte Carlo generator should yield both the momenta of produced particles and
the space-time coordinates of their creation (or last interaction) points.  Even
if we avoid troubles with the uncertainty principle by using the Wigner function
approach, such a generator seems reliable only for heavy ion collisions.  It has
been constructed also for the $e^+ e^-$ collisions~\cite{eg}, but localizing the
hadron creation point in the parton-based Monte Carlo program for lepton and/or
hadron collisions is a rather arbitrary procedure, and it is hard to say what
does one really test comparing such a model with data.
\par
It seems to be the
best procedure to take into account the interference effects before generating
events.  Unfortunately, this was done till now only for the JETSET generator for
a single Lund string~\cite{ar,ar2,tnr,hr}, and a generalization for multi-string
processes is not obvious.  No similar modifications were yet proposed for other
generators.
\par
The most popular approach, applied since years to
the description of BE effect in various processes, is to shift the final state
momenta of events generated by the PYTHIA/\\JETSET generators~\cite{sjo,ls}.  The
prescription for a shift is such as to reproduce the experimentallly observed
enhancement in the ratio
\begin{equation} 
c_2(Q)=\frac{<n>^2}{<n(n-1)>}
\frac{\int d^3 p_1 d^3 p_2 \rho_2 (p_1,p_2) \delta (Q-\sqrt{-(p_1-p_2)^2})}
{\int d^3 p_1 d^3 p_2 \rho_1 (p_1) \rho_1 (p_2) \delta (Q-\sqrt
{-(p_1-p_2)^2})}, 
\label{eq:ratio}
\end{equation}
\noindent which is a function
of a single invariant variable $Q$. The value of this function is close to one for the
default JETSET/PYTHIA generator. One parametrizes often this ratio by
\begin{equation} c_2(Q)=1+\lambda exp(-R^2 Q^2), \label{eq:factor}
\end{equation}
\noindent where $R$ and $\lambda$ are parameters interpreted as
the source radius and "incoherence strength", respectively.
\par
After performing the shifts, all the CM 3-momenta of final state particles are
rescaled to restore the original energy.  In more recent versions of the
procedure~\cite{ls2} "local rescaling" is used instead of the global one.  In
any case, each event is modified and the resulting generated sample exhibits now
the "BE enhancement":  the ratio ~(\ref{eq:ratio}) is no longer close to one,
and may be parametrized as in ~(\ref{eq:factor}).
\par
There is no theoretical justification for this procedure, so it should be regarded as
an
imitation rather than implementation of the BE effect.  Its success or failure in
describing data
is the only relevant feature.  Unfortunately, whereas the method is very useful
for the description of two-particle inclusive spectra, it fails to reproduce
(with the same fit parameters $R$ and $\lambda$) the three-particle
spectra~\cite{ua1} and the semi-inclusive data~\cite{na22}.  This could be
certainly cured, e.g., by modifying the shifting procedure and fitting the
parameters separately for each semi-inclusive sample of data.  However, the
fitted values of parameters needed in the input factor ~(\ref{eq:factor}) used
to calculate shifts are quite different from the values one would get fitting
the resulting ratio ~(\ref{eq:ratio}) to the same form~\cite{fw1}. This
was shown recently in a much more detailed study~\cite{sm}.  Thus it seems to be very
difficult to learn something reliable on the space-time structure of the source
from the values of fit parameters in this procedure.
\par
All this has led to the revival of weight methods, known for quite a long
time~\cite{pra},
but
plagued with many practical problems.  The method is clearly justified within
the formalism of the Wigner functions, which allows to represent (after some
simplifying assumptions) any distribution with the BE effect built in as a
product of the original distribution (without the BE effect) and the weight
factor, depending on the final state momenta~\cite{bk}.  With an extra
assumption of factorization in momentum space, we may write the weight factor
for the final state with $n$ identical bosons as

\begin{equation} W(p_1,...p_n)=\sum
\prod_{i=1}^n w_2(p_i,p_{P(i)}),
\label{eq:weight}
\end{equation}

\noindent
where the sum extends over all permutations ${P_n(i)}$ of $n$ elements, and
$w_2(p_i,p_k)$ is a two-particle weight factor reflecting the effective source
size.  A commonly used simple parametrization of this factor for a Lorentz
symmetric source is

\begin{equation} w_2(p,q)=exp[(p-q)^2/2Q_0^2],
\label{eq:wf}
\end{equation}

The only free parameter is now $Q_0$, representing the inverse of the effective
source size.  In fact, the full weight given to each event should be a product
of factors ~(\ref{eq:weight}) calculated for all kinds of bosons; in practice,
pions of all signs should be taken into account.  Only direct pions and the
decay products of $\rho, K^*$ and $\Delta$ should be taken into account, since for
other
pairs much bigger
$R$ should be used, resulting in negligible contributions.

\par
The main problem of the weight methods is
that weights do change not only the Bose - Einstein ratio ~(\ref{eq:ratio}),
 but also many other distributions.  Thus with the
default values of free parameters (fitted to the data without weights) we
find inevitably some discrepancies with data after introducing weights.
\par
We
want to make clear that this cannot be taken as a flaw of the weight method.
There is no measurable world ``without the BE effect'', and it makes not much
sense to ask, if this effect changes e.g.  the multiplicity distributions.  If
any model is compared to the data without taking the BE effect into account, the
fitted values of its free parameters are simply not correct.  They should be
refitted with weights, and then the weights recalculated in an iterative
procedure.  This, however, may be a rather tedious task.
\par
Therefore we use
a simple rescaling method proposed by Jadach and Zalewski ~\cite{jz}.  Instead
of refitting the free parameters of the MC generator, we rescale the BE weights
(calculated according to the procedure outlined above) with a simple factor
$cV^n$, where $n$ is the global multiplicity of "direct" pions, and $c$ and $V$
are fit parameters.  Their values are fitted to minimize
\begin{equation}
\chi^2 = \sum_n[cV^nN^w(n) - N^0(n)]^2/N^0(n)
\label{eq:chi2}
\end{equation}
where $N^0(n)$ is the number of events for the multiplicity $n$ without weights,
and $N^w(n)$ is the weighted number of events.  This rescaling restores the
original multiplicity distribution ~\cite{fww}.  In addition, the single
longitudinal and transverse momentum spectra are also restored by this rescaling
~\cite{fww}.
\par
Obviously, for a more detailed analysis of the final states,
single rescaling may be not enough.  E.g., since different parameters govern the
average number of jets and the average multiplicity of a single jet, both should
be rescaled separately to avoid discrepancy with data.  Let us stress once again
that such problems arise due to the use of generators with improperly fitted
free parameters, and do not suggest any flaw of the weight method.  Another
problem is that our formula for weights~(\ref{eq:weight}) is derived using some
approximations, which are rather difficult to control~\cite{bk}.  We can justify
them only {\it a posteriori} from the phenomenological successes of the weight
method.
\par
Last but not least, the main practical difficulty with formula (\ref{eq:weight}) is
the
factorial increase of the number of terms in the sum with increasing multiplicity of
identical pions $n$.  For high energies, when $n$ often exceeds 20, a straightforward
application of formula (\ref{eq:weight}) is impractical~\cite{hay}, and some
authors ~\cite{jz,kkm} replaced it with simpler expresssions, motivated by some
models.  It is, however, rather difficult to estimate their reliability.
\par
We have recently proposed two ways of dealing with this problem.  One method
consists of a truncation of the sum (3) up to terms, for which the permutation
$P(i)$ moves no more than 5 particles from their places ~\cite{fw2}. However, it is
difficult to claim a priori that such a truncation does not change the results which
would
be obtained using the full series (2).
\par
Therefore a second way of an approximate calculation of the
sum (2) was proposed ~\cite{jw}.  Since this sum, called a permanent of a matrix
built from weight factors $w_{i,k}$, is quite familiar in field theory, one may
use a known integral representation and approximate the integral by the saddle
point method. However, this method is reliable only if in each row (and column)
of the matrix there is at least one non-diagonal element significantly different
from zero.  Thus the
prescription should not be applied to the full events, but to the clusters, in
which each momentum is not far from at least one other momentum.  The full
weight is then a product of weights calculated for clusters, in which the full
event is divided.
\par
The considerations presented above suggested the necessity of combining these
two methods.  After dividing the
final state momenta of identical particles into clusters, we used for small clusters
exact
formulae presented in~\cite{fw1,fw2}.  For large clusters (with more than five
particles)
we compared two approximations (truncated series and the integral representation) to
estimate their reliability and the sensitivity of the final results to the
method.  Obviously, the results depend also on the clustering algorithm:  if we
restrict each cluster to particles very close in momentum space, the neglected
contributions to the sum ~(\ref{eq:weight}) from permutations exchanging pions
from different clusters may be non-negligible, and if the cluster definition is
very loose, the saddle point approximation may be unreliable.  This was then
also checked to optimize the algorithm used.  We found that the truncated series
approximation was sufficient in all cases we checked on~\cite{fww}. Our method was 
already applied to the analysis of $W^+W^-$ production ~\cite{jww}.

\section{The data and their analysis}
Extremely high
statistics collected in LEP for the hadronic decays of $Z^0$-s produced in
   the $e^+e^-$ collisions allowed to investigate in detail many interesting
effects. In particular, the interference effects due to the Bose-Einstein
statistics for pions were analyzed in several experiments.  \par The
notorious difficulty in measuring the Bose-Einstein interference effects
(BE effects) in the multiparticle production is the proper choice of
   reference sample. In the early investigations the ratio of numbers of
"like-to-unlike" charged pion pairs was analyzed as a function of three-
or four-momentum difference squared
\begin{equation}R_{BE}(Q^2)=\frac{\int\rho_2^{++}(p_1,p_2)\delta[(p_1-p_2)^2+Q^2]d^3p_1
d^3p_2}{\int\rho_2^{+-}(p_1,p_2)\delta[(p_1-p2)^2+Q^2]d^3p_1d^3p_2}
\label{eq:6}
\end{equation}
Obviously, the
denominator has Breit-Wigner peaks around the values of $Q^2$ corresponding
to masses of resonances in the $\pi^+\pi^-$system (and other peaks due to
the maxima in mass spectra from 3-pion resonances) which are absent in the
numerator.  To estimate properly the BE effects one should subtract these
maxima, or to exclude the "resonance regions" from the $Q^2$ range used in
fitting $R_{BE}$ to the chosen function.
\par
Therefore recently a more common choice
for the denominator was the "uncorrelated background" formed e.g. by
   choosing pairs of "like sign" pions from different events, which led to
   the definition presented in the former section (\ref{eq:ratio}).
    Here the main problem comes from neglecting the energy-momentum conservation
effects (present in the numerator and absent in the denominator). However,
for high energies and restricted range of momenta (e.g. the "central
region", often used for the analysis) such effects are expected to be
rather small.
\par
This method is easy to apply for hadronic or heavy ion
    collisions, where the initial momenta form the natural symmetry axis
in the CM frame. For $Z^0$ decays the typical events are not aligned with
the momenta of the initial $e^+e^-$ pair. Thus the momenta chosen from
different events should be rotated to the same symmetry axis before
calculating $Q^2$.  Unfortunately, such a procedure is not well-defined:
using sphericity, thrust or other variables one obtains different values
of rotation angles.  Moreover, the three- and many-jet events do not have
well-defined symmetry axis, and limiting the analysis to the two-jet
events would be rather arbitrary (and dependent on the jet definition).

\vspace{1cm}

\epsfxsize=10cm

~~~~~~~~~~~~~~~\epsfbox{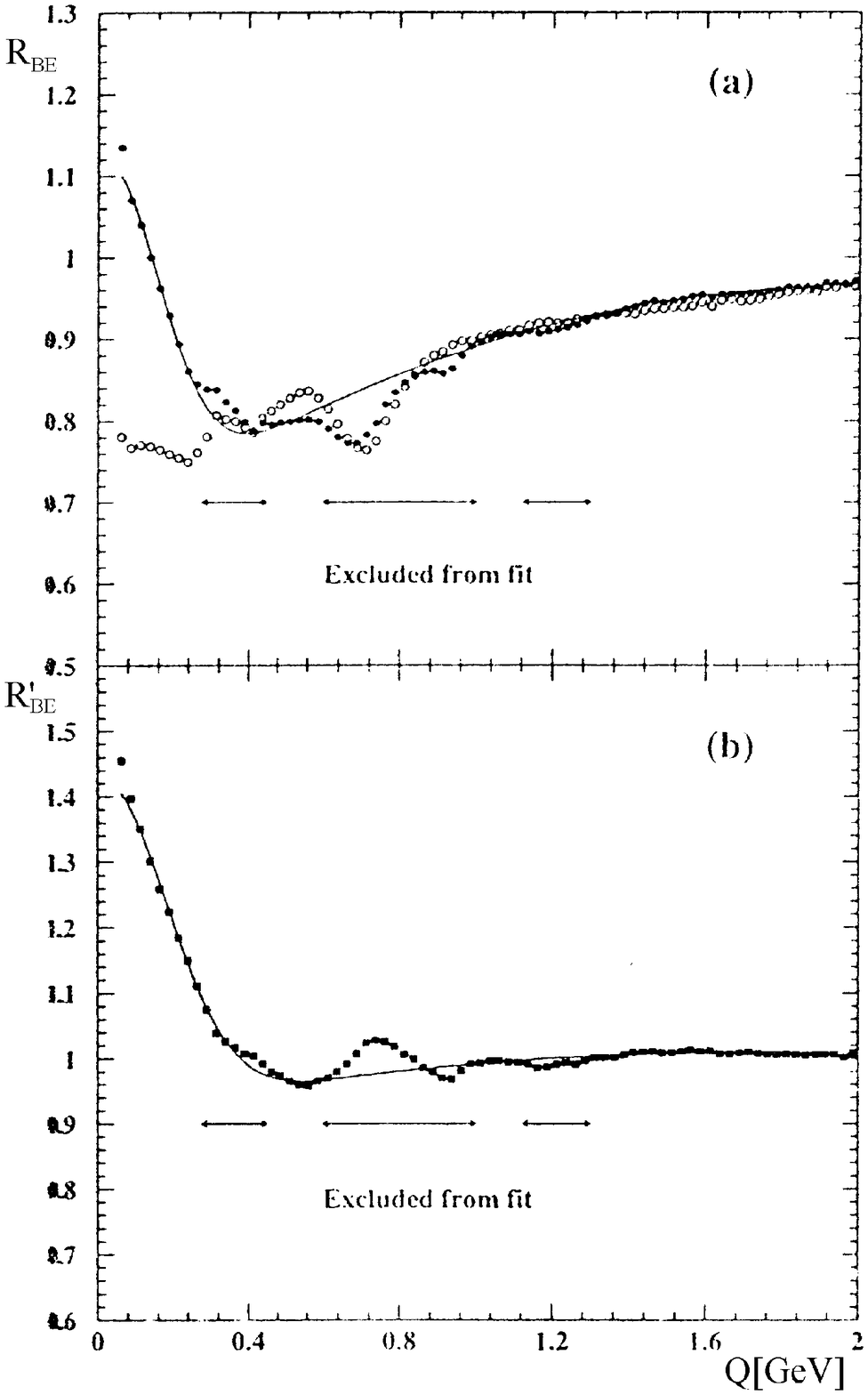}

\par
 {\bf Fig.1.} {\sl  a)  The BE ratio $R_{BE}$ (6) from the OPAL data (black points) and 
 from the JETSET MC 
without BE effect (open points); b) the double ratio $R'_{BE}$ (7) from 
the same data. Solid lines are fits to the form (8) [25].}

\par
Therefore the best strategy seems to be using Monte Carlo generators,
   which are rather reliable for the process under discussion. Instead of
analyzing the BE ratio $R_{BE}$ one considers the "double ratio", i.e. the
ratio of $R_{BE}^{exp}$ given by the data and $R_{BE}^{MC}$ computed
from the MC generated events
\begin{equation}
R'_{BE} = R_{BE}^{exp}/ R_{BE}^{MC}.
\label{eq:7}
\end{equation}

\par 

Analogous "double ratio" has been defined also for the first definition of the "BE ratio"
(\ref{eq:ratio})~\cite{aleph0}. The results obtained for two double ratios are 
inconsistent.
In the following we use only the definition (\ref{eq:7}), preferred recently by the 
experimental 
groups~\cite{opal,aleph}.
\par
 The data with largest statistics has been presented by the OPAL group
both for $R'_{BE}$ and for $R_{BE}$. They compared the parameter values from
the fits (with resonance regions excluded) to the function
\begin{equation}
f(x) = \kappa[1+\lambda exp(-R^2x^2)](1+\alpha x^2) .
\label{eq:8}
\end{equation}
The data
(based on 3.6 milion of events) were shown for $x=Q$ and for 
$x= q_t=\sqrt{({\bf p_1}_{t} 
- {\bf p_2}_{t})^2}$
~\cite{opal}.
We remind here in Fig.1 the data and fits for $R_{BE}(Q)$ and
$R'_{BE}(Q)$. The values of $\lambda$ and R differ quite significantly, as shown in Table 
I.

\begin{table}[ht]
{\small
\begin{center}
TABLE I. Fits to OPAL data.
\end{center}
}
\begin{center}{
\begin{tabular}{lccc}
\hline
\hline
 \w & $R(fm)$ & $\lambda$ & $\chi^2/d.o.f.$ \\
\hline
Reference fit to $R_{BE}$\w & $0.955 \pm 0.012$ & $0.672 \pm 0.013$ &
402/40 \\
Fit with Q binning = 50 MeV \w & $0.962 \pm 0.013$ & $0.667 \pm 0.014$ &
204/17\\
Fit to $R'_{BE}$ \w & $0.793 \pm 0.015$ & $0.577
\pm 0.010$ & 185/40\\
\end{tabular}}
\end{center}
\end{table}

The other important conclusion drawn
by the authors of that work was that even the fit for the double ratio
requires cutting off the resonance regions.  This is because the standard
JETSET/PYTHIA MC generator does not describe satisfactorily the ratio $R_{BE}$
not only for small Q (where the BE effect may appear), but also in the
resonance region.
\par
Let us add that one may interpret this
disagreement as a signal that the production of $K_S^0$ and $\rho$ is
overestimated in this MC:  the dips in experimental $R_{BE}$ at Q = 0.40 and
 0.72 GeV (corresponding to $K_S^0$ and $\rho$ masses, respectively) are less
sharp in data than in MC, which produces bumps in $R'_{BE}$.
\par
In any
case, the OPAL data show that one must be careful in interpreting the
shape of the BE ratio (or double ratio) when using the unlike sign pairs as
the reference sample.  The fit parameters, which are tentatively
interpreted as characterizing the source incoherence and geometrical size,
depend significantly on the method of analysis.  Moreover, the imperfect
description of resonance production in MC may distort the shape of double
ratio (2), making the detailed success or failure of the fits to specific
functional forms rather ambiguous.
\par
This seems to be the case for the
recently presented ALEPH data for $Z^0$ decay used in the analysis of W-pair decays
~\cite{aleph}.
As shown in Fig. 2, the data for $R'_{BE}$ deviate from a smooth fit
in a very similar way as the OPAL data shown in Fig. 1b (if one excludes
the resonance ranges, the fit would go through points for Q between 0.4
and 0.6 GeV leaving the points below 0.4 GeV above the curve, and the
excess around 0.7 GeV would be even more pronounced). Due to smaller
 statistics (below 100 000 events) much wider bins (of 0.1 GeV) were chosen.

\vspace{1cm}

\epsfxsize=10cm

~~~~~~~~~~~~~~~\epsfbox{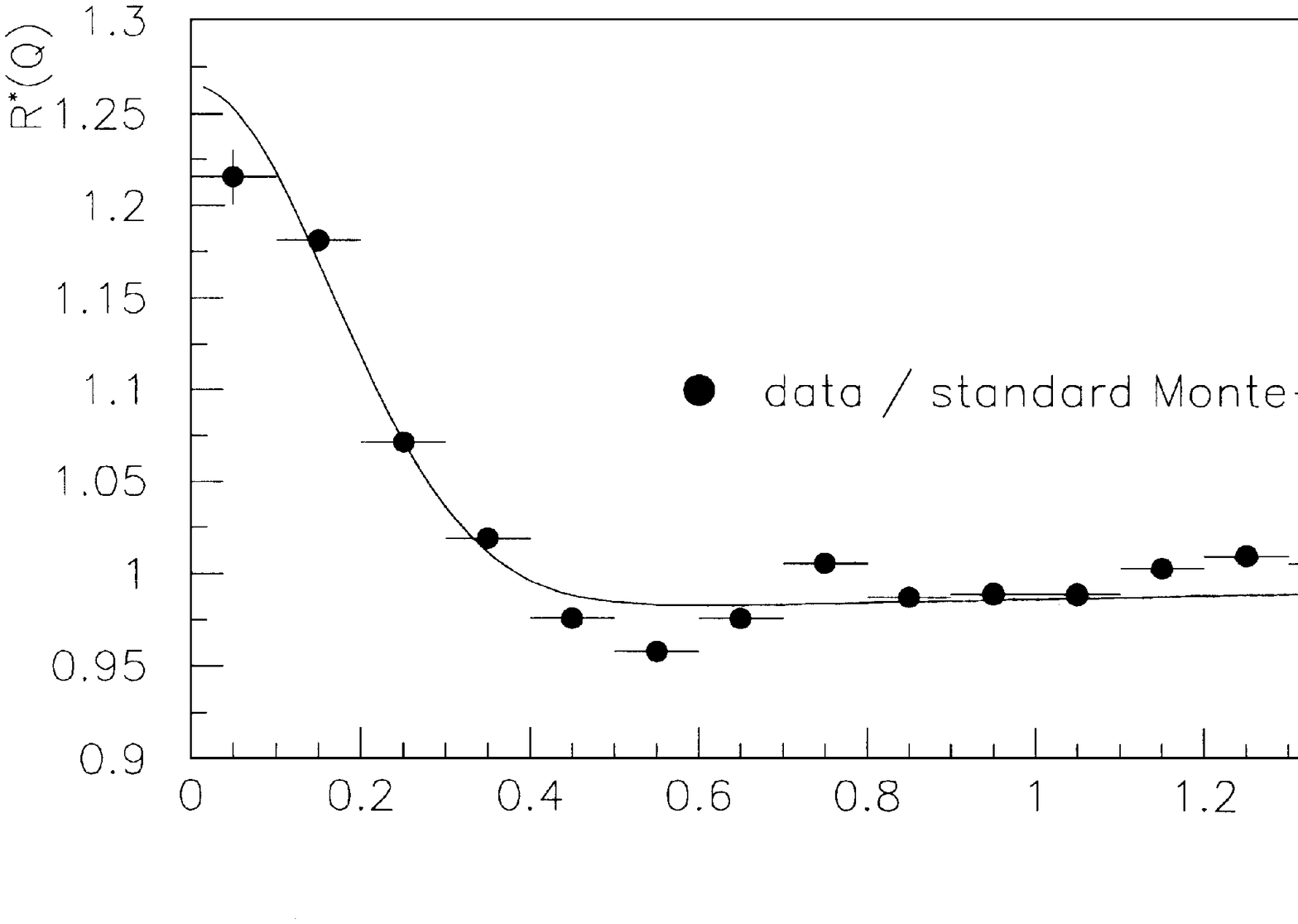}

\par
 {\bf Fig.2.} {\sl The BE double ratio $R'_{BE}$ (7) from the ALEPH data. The 
 solid line is a fit to the form (8)~\cite{aleph}.}

\par
Nevertheless, it seems that the
claim of the authors of Ref.~\cite{aleph} that the data show a clear
preference for
one of the three models implementing the BE effect into MC requires more
careful analysis.  In fact, if one would cut off the resonance regions (as
done for OPAL data) certainly no conclusions could be drawn. If the full
range of Q is used, one should ask before comparing models  if they
influence the size and/or shape of the resonance contributions to the BE
ratios. We will come back later to these questions.

\section{ Weight methods - assumptions and results}
\par
Obviously, the Gaussian two-particle weight factor used in our method
is just a simple ansatz. There are no deep reasons to expect such a shape
(corresponding to a Gaussian distribution of two-particle birthplace
position differences) for the weight factor. In fact, the Wigner functions (which serve 
to
define the weight factors \cite{bk}) are not even necessarily positive.  Thus
it is interesting not just to fit the Gaussian width to the data, but to
investigate generally how the shape of $R_{BE}$ (or $R'_{BE}$)
depends on the parameters of the two-particle weight factors for their
different functional forms. In principle the data may allow to find the
proper factor in momentum space and to deduce from that some information
on the space-time structure of the source.  For simplicity, we restrict
ourselves here to the analysis of data presented in Section 3, and thus to
   the discussion of functions of a single variable $Q^2$ (which
corresponds to the assumption of a space-time symmetric source).
\par
We have performed MC generation of hadronically decaying $Z^0$ -s produced
   in the $e^+e^-$ collisions at the peak energy with the standard PYTHIA/JETSET
   generator.
   Each event was assigned the weight calculated according to our method
   described in Section 2 with the two-particle weight factor given alternatively
   by a Gaussian (\ref{eq:wf}), exponent ($e^{-Q/Q_0}$), step function (1 for $Q \le 
Q_0$,
   0 for $Q > Q_0 $), diffuse
   step function ($2/(1+e^{Q/Q_0})$), double Gauss
   ($\alpha e^{-Q^2/2Q_0^2}+(1-\alpha)e^{-Q^2/Q_0^2}$) or an oscillating exponent
($e^{-Q/Q_0}sin(\alpha Q)/\alpha Q$) with various
   values of parameters for each
   form of the weight factor.
   For each set of parameters one million of events was generated.
\par
   For all samples the double ratio  $R'_{BE}$ was calculated for twenty equal
   bins in Q in the
   range $0<Q<2$ GeV. In all cases we obtained for small Q the values of $R'_{BE}$
   significantly above 1, and for $Q>>Q_0$ the values compatible with 1. 
An example (for exponential weight factor with $Q_0 = 0.316$) is shown in Fig 3.

\vspace{0.5cm}

\epsfxsize=10cm

~~~~~~~~~~~~~~~\epsfbox{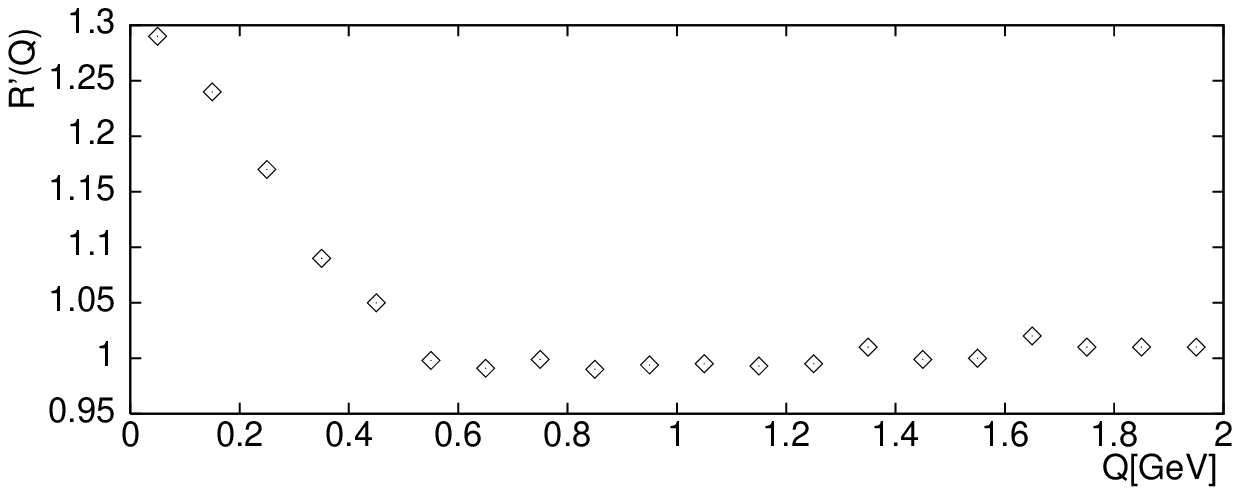}

\vspace{0.5cm}
\par
 {\bf Fig.3.} {\sl The BE double ratio $R'_{BE}$ (7) for exponential weight factor with 
$Q_0 = 0.316$.}
\vspace{0.5cm}
   
\par
   A reasonable
   fit to the form (2) for the first few bins in Q was obtained in all the cases.
   We summarize all the results
   in Table II and in a simple two-dimensional diagram (Fig. 4), in which the fitted
values of $\lambda$ and R are given.

\vspace{1cm}

\epsfxsize=10cm

~~~~~~~~~~~~~~~\epsfbox{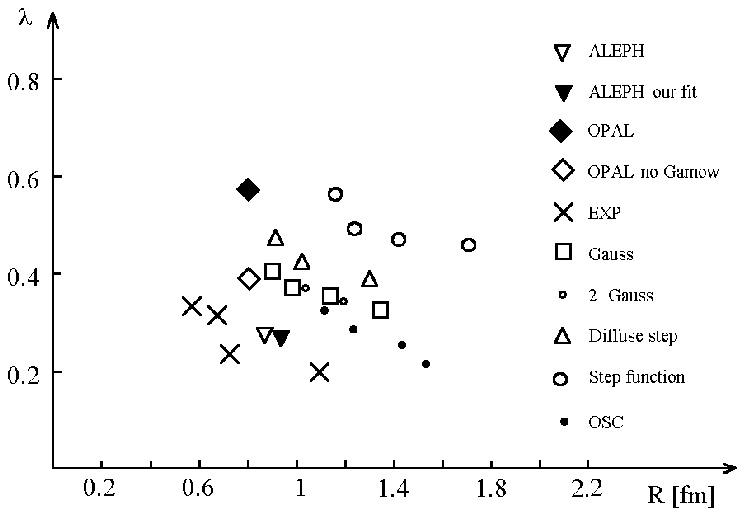}

\par
 {\bf Fig.4.} {\sl The parameters of the fits to the form (2) for first six bins of 
$R'_{BE}$  
from the 
samples of events generated with various weights factors: step function (circles), 
diffuse 
step (triangles), Gaussian (squares), double Gaussian (open points), exponent with 
oscillations (black points) and exponent (crosses). Parameters of the fits to the data 
are also shown as black diamond (OPAL~\cite{opal}), open diamond (our fit to OPAL without 
Gamow 
correction), open triangle (ALEPH~\cite{aleph}) and black triangle (our fit to ALEPH).}\\

\par
Each symbol in the diagram corresponds to a given form
   of the weight factor. Increasing values of $Q_0$ correspond always to the increasing
   values of  $\lambda$ and decreasing values of R. The values fitted by the
OPAL and ALEPH
collaborations to their data are also shown. Since these fits are performed to
   more complex formulae (e.g. (\ref{eq:8})) and in different ranges of Q, we have 
checked
   that very similar values result from our simple fits.  We will comment later 
   on other LEPII experiments.

\begin{table}[ht]
{\small
\begin{center}
TABLE II. Fits to MC results.
\end{center}
}
\begin{center}{
\begin{tabular}{lccccc}
\hline
\hline
 Factor\w & $Q_0[GeV]$ & $\alpha$ & $\lambda$ & $R[fm]$ & $R/R_{eff}$\\
\hline
Step function \w & 0.141   &       & 0.46 & 1.73 &   0.61\\
              \w & 0.158   &       & 0.47 & 1.44  &  0.57 \\
              \w & 0.173   &       & 0.49 & 1.23  &  0.53\\
              \w & 0.187   &       & 0.56 & 1.16  &  0.54\\
Exponent      \w & 0.2    &       & 0.20 & 1.10   &  1.10\\
              \w & 0.244   &       & 0.24 & 0.73  &  0.90\\
              \w & 0.283   &       & 0.31 & 0.66  &  0.94\\
              \w & 0.316  &       & 0.33 & 0.57   &  0.90\\
Gauss         \w & 0.2   &       & 0.33 & 1.34    &  0.75\\
              \w & 0.224   &       & 0.36 & 1.15  &  0.72\\
              \w & 0.245   &       & 0.37 & 0.99  &  0.68\\
              \w & 0.265   &       & 0.41 & 0.89  &  0.66\\ 
Diffuse step  \w & 0.172   &       & 0.39 & 1.30   & 0.73\\ 
              \w & 0.2     &       & 0.42 &  1.02  &   0.66\\ 
              \w & 0.224   &       & 0.47 &  0.90  &  0.65\\ 
2 Gauss       \w & 0.2   & 0.2   & 0.34 & 1.19    & 0.74\\
              \w & 0.2   & 0.5   & 0.37 & 1.04    & 0.73\\
Oscill. exp.  \w & 0.244 & $8\ GeV^{-1}$ & 0.21 & 1.53 &  0.69   \\
              \w & 0.316  &       & 0.25 & 1.43        & 0.65\\
              \w & 0.448  &       & 0.28 & 1.23        &  0.55\\
              \w & 0.632   &       & 0.32 & 1.12       & 0.49\\

\end{tabular}}
\end{center}
\end{table}

\par
$Q_0$ is not a best measure of "width in momentum space" when one compares different 
shapes of the weight factor. It seems more natural to use $\overline Q$, which is just 
the 
average value of $Q$ for a given weight factor
\begin{equation}
\overline Q = \frac{\int_0^{\infty}Qw_2(Q)dQ}{\int_0^{\infty}w_2(Q)dQ}.
\label{eq:aver}
\end{equation}
If we define $R_{eff} = 1/\overline Q$ as the "effective source size" we find a close 
relation between the fitted values of $R$ and the input values of $R_{eff}$. The ratio 
$R/R_{eff}$, quoted in the last column of Table II, is in the range $0.5 - 1$ for almost 
all values of $Q_0$ and all shapes of the weight factor.   

\par
   The first obvious conclusion from Fig. 4 is that the fits to the two experimental sets 
of data
are strongly different. This is partly due to the Gamow correction for the effect of
   electromagnetic final state interactions, which was used by OPAL and neglected
   by ALEPH. Since it has been recently argued that this correction badly
   overestimates the effect ~\cite{mp}, we have shown also an estimate for the OPAL point
   without Gamow
   correction (obtained using the values for correction quoted in Ref. ~\cite{opal}).  
Now
the
   two data points are much closer, but still they differ significantly.
\par
   The values resulting from our generated samples cover quite a wide range both in
   $\lambda$ and R. Certainly it is possible to reproduce any of the
   experimental values by a
   suitable
   choice of the form and free parameters of our weight factors. However, we do not
   think it would give a valuable information on the "proper" choice of the
   weight factor reflecting the "real" space-time structure of the source.
\par
   There are a few reasons for such scepticism. First, as noted in the previous section,
   the shape of "experimental" $R'_{BE}$ strongly depends on the quality of "reference 
MC". 
Since we know that already at $Q \cong 0.40
GeV$ the influence of (poorly described) $K_S^0,\ \eta$ and $\eta '$ contributions 
distorts the shape,
   the fitted values of $R$ around $ 1fm \cong 0.2 GeV^{-1}$ are not very
   reliable. The situation is even less clear if one fits wider range of Q using a more
elaborate ansatz for $R'_{BE}$ (e.g. (\ref{eq:8})). Here the choice of "cut out resonance
region" may influence quite significantly the fit results. Moreover, the method of
implementing
   the BE effect may influence the resonance contribution. For our method the absence of
peaks or bumps in the double ratio for "MC with weights" over "MC without weights"
(as seen, e.g., in Fig. 3)   suggests
   that this is not the case (the oscillations shown in Fig. 2) of Ref.~\cite{aleph} 
    may result from lower
statistics). This should be also checked for any other method.
\par
   Second, the choice of "direct" pions affected by the BE effects described in Section 2
(although the same in our method and in the Sjoestrand's momentum shift method) is by no
means unique. In fact, even for the fastest-decaying resonances the effective "source
radius" for the decay products should not be exactly the same as for the "real direct
pions". Thus the choice of a single value of $Q_0$ in our weight factor is certainly an
oversimplification. The total exclusion of the decay products of long-living
   resonances is also an approximation: some of them may be born quite close to the
collision point and contribute to the visible BE effect. The selection and treatment of
"direct" pions included in calculating weights (or subject to the momentum shift in the
Sj\"oestrand's method) influences strongly the resulting value of $\lambda$.
\par
   Third, the standard fitting method of BE ratios or double ratios is also a subject of
criticism ~\cite{dav}. More plausible methods suggest bigger uncertainties of the fit
parameters.
\par
   The situation would not improve if we add the other published data, which used 
   much lower statistics~\cite{delphi1} or different form of the fitted function
   ~\cite{delphi2}.  One may
shortly summarize that it is too early to deduce details of the source space-time
   structure from the existing data, even if we assume the applicability of all the
assumptions used in formulating the weight method.
\par
    Another interesting feature is the relative stability of our results with respect
    to the detailed shape of two-particle weight factor. The fitted values of $\lambda$
     fall
in the range of values 0.2 - 0.5 for all the functions considered. They are highest
for the
step function, lower for "diffuse step function", still lower for a Gaussian and
    lowest for the exponent. One may summarize that for the same "half width" of the
    weight factor function one gets higher value of $\lambda$ if the function stays 
longer
near the value of  1
at small Q. As already noted, $\lambda$ depends much stronger on the percentage of pions
counted as "direct". The values of $\lambda$ below 1 reflect mainly the percentage of 
direct 
like-sign pion pairs, which was already observed when using Sj\"ostrand's method of BE 
effect implementation~\cite{delphi2}.
\par
   A new result is the observed relative insensitivity of the results on the
oscillations in the two-particle weight factor. In fact, the oscillation half-period
serves as an effective range of correlations if it is smaller than $Q_0$. 
Therefore $R_{eff}$ is almost the same for all the values of $Q_0$ quoted in the Table 
II.
The non-positive
oscillating values of weight factor for larger Q do not result in the values of $R'_{BE}$
oscillating around 1, as one could naively expect; averaging over many pairs in each
   event and over many events kills any trace of oscillations. This is quite important,
   as many simple forms of weight factor in space-time (e.g. the step function) results 
in
   oscillating form in momentum space.
\par
   On the other hand, there is an obvious dependence of fitted values of R on the
   parameter $Q_0$ (or $R_{eff}$), which determine how fast the weight factor decreases 
with increasing Q. $R/R_{eff}$ has quite similar values for various shapes of the weight 
factors and various values of $Q_0$.
Thus in our method R reflects in some sense the "source size", as expected. As already
noted, this is true also for the oscillations in momentum space.  Increasing $Q_0$ one
increases also
   fitted values of $\lambda$, but this effect is weaker.

\section{Conclusions and outlook}

   We have analyzed the data on  BE effect in multihadron states obtained 
   from $Z^0$ decays within the framework of the weight method of implementing 
   the effect. We show that the resonance contributions are not satisfactorily 
   described by the standard MC generators, which makes the analysis of 
   "double ratio" data quite difficult. In particular, the choice of the 
   "proper" two particle weight factor which would describe the data best 
   is obviously dependent on the way the data are processed. Thus one should 
   improve (if possible) the quality of MC generators before drawing any 
   serious conclusion on the agreement or disagreement between the models 
   and data on the BE effect. The alternative is to elaborate better methods 
   of preparing the "reference sample" using only the like-sign pairs, which 
   are much less affected by the resonance effects.
\par
   On the other hand, we found that the analysis makes no great difference 
   between the positively defined and oscillating weight factors in momentum 
   space, provided they correspond to  similar "half-width". This means that 
   the results are not too strongly dependent on the "sharpness of the source 
   boundary" in space-time; one may hope to recover the proper "source size" 
   in our method.
\par 
   Let us stress that the analysis presented in this paper is just the first 
   step: we do not discuss the anisotropy of source nor the time dependence 
   reflected in the BE effect. Also, we do not consider here the possibilities 
   of comparing the effect for selected classes of events (defined i.e. by 
   multiplicity, by event shape parameters or by some special triggers). In 
   our opinion, the weight method is well suited to discuss these subjects 
   and we hope to consider them in near future.
\section{Acknowledgements}
The authors are grateful to A. Bia\l{}as for useful remarks. The financial 
support from KBN grants No 2 P03B 086 14 and No 2 P03B 010 15 are acknowledged. A. 
Gagatek
provided patient assistance with figure preparations.


\begin{thebibliography}{99}
\bibitem{hbt}R.  Hanbury-Brown and R.Q.  Twiss,
 {\it Nature} {\bf 178} {(1956)} {1046}.
%\bibitem{web}B.R.  Webber, {{\it J.  Phys.}  G} {\bf 24} {(1998)} {287}.
\bibitem{bgj}D.H.  Boal, C.-K.  Gelbke and B.K.  Jennings,  {\it Rev.
Mod.  Phys.} {\bf 62} {(1990)} {553}; R.M. Weiner, e-print hep-ph/9904389.
\bibitem{sul}J.P.  Sullivan {\it et al.}, {\it Phys. Rev. Lett.} {\bf 70} {(1993)}
{3000}.
\bibitem{zha}Q.H.  Zhang {\it et al.}, {{\it Phys. Lett.} B} {\bf 407} {(1997)} {33}.
\bibitem{eg}J.  Ellis and K.  Geiger, {\it Phys. Rev.} {D} {\bf 54}
{(1996)} {1967}; K. Geiger, J. Ellis, U. Heinz and U.A. Wiedemann,
e-print hep-ph/9811270.
\bibitem{ar}B.  Andersson and M.  Ringn\`er, {\it Nucl.
Phys.} {B} {\bf 513} {(1998)} {627}.
\bibitem{ar2}B.  Andersson and M.
Ringn\`er, {\it Phys. Lett.} {B} {\bf 421} {(1998)} {283}.
\bibitem{tnr}\u{S}.  Todorova-Nov\`a and J.  Rame\u{s}, e-print hep-ph/9710280.
\bibitem{hr}J.  H\"akkinen and M.  Ringn\`er, {{\it Eur. Phys. J.} C} {\bf 5} {(1998)
} {275}.
\bibitem{sjo}T.  Sj\"ostrand and M.  Bengtsson, {\it Comp.  Phys.
 Comm.} {\bf 43} {(1987)} {367}; T.  Sj\"ostrand, {\it Comp.  Phys.  Comm.}
 {\bf 82} {(1994)} {74}.
\bibitem{ls}L.  L\"onnblad and T.  Sj\"ostrand, {\it Phys. Lett.} {B} {\bf 351}
{(1995)} {293}.
\bibitem{ls2}L.  L\"onnblad and T.  Sj\"ostrand, {{\it Eur.  Phys.  J.}
C} {\bf 2} {(1998)} {165}.
\bibitem{ua1}Y.F.  Wu {\it et al.}, in {\it Proc.  Cracow Workshop on
Multiparticle Production "Soft Physics and Fluctuations"}, ed.  A.  Bia{\l}as {\it
et al.}, p.22, World Sci., Singapore 1994.
\bibitem{na22}N.M.  Agababyan {\it et al.}, {\it Phys. Lett.} {B} {\bf 332} {(1994)}
{458}.
\bibitem{fw1}K.  Fia{\l}kowski and R.  Wit, {\it Z. Phys.} {C} {\bf 74} {(1997)}
{145}.
\bibitem{sm}O.  Smirnova, B.  L\"orstad, R.  Mure\c{s}an, Proceedings of the
8th
Int. Workshop on Multiparticle Prod., {\it Correlations and Fluctuations '98}
(M\'atrah\'aza, Hungary), ed. T. Cs\"org\H{o}, S. Hegyi, G. Jancs\'o and R.C. Hwa,
 p.20, World Sci. Singapore 1999.
\bibitem{pra}S.  Pratt, {\it Phys. Rev. Lett.} {\bf 53} {(1984)} {1219}.
\bibitem{bk}A.  Bia{\l}as and A.  Krzywicki, {\it Phys. Lett.} {B} {\bf 354} {(1995)}
{134}.
\bibitem{hay}S.  Haywood, Rutherford Lab.  Report RAL-94-074.
\bibitem{jz}S.  Jadach and K.  Zalewski,  {{\it Acta Phys.  Pol.}
B} {\bf 28} {(1997)} {1363}.
\bibitem{kkm}V.  Kartvelishvili, R.  Kvatadze and R.  M{\o}ller,
{\it Phys. Lett.} {B} {\bf 408} {(1997)} {331}.
\bibitem{fw2}K.  Fia{\l}kowski and R.  Wit, {{\it Eur.  Phys.  J.}
C} {\bf 2} {(1998)} {691}.
\bibitem{jw}J.  Wosiek, {\it Phys. Lett.} {B} {\bf 399} {(1997)} {130}.
\bibitem{fww}K.  Fia{\l}kowski, R.  Wit and J.  Wosiek, {{\it Phys.  Rev.} D}
 {\bf  57} {(1998)} {094013}.
\bibitem{jww}K.  Fia{\l}kowski and R.  Wit,  {\it Phys.  Lett.} {B} {\bf 438} {(1998)}
{154}. 
\bibitem{aleph0} ALEPH collaboration, D. Decamp {\it et al.}{\it Z. Phys.} {C}
{\bf 54} {(1992)} {75}.
\bibitem{opal}OPAL Collaboration, G. Alexander {\it et al.} {\it Z. Phys.} {C}
{\bf 72} {(1996)} {389}.
\bibitem{aleph}The ALEPH Collaboration, Collaboration report ALEPH 99-027,
contribution CONF 99-021 to the 1999 winter conferences.
\bibitem{mp}M.G. Bowler, {\it Phys.  Lett.} {B} {\bf 270} {(1991)} {69}; M.Biyajima {\it 
et al.} {\it Phys.  Lett.} {B} {\bf 353} {(1995)} {340}; G.Baym and P. Braun-Munzinger, 
{\it Nucl. Phys.} {A} {\bf 610} {(1996)} {286c}; H. Merlitz and D. Pelte, {\it Phys. 
Lett.} {B} {\bf 415} {(1997)} {411}.
\bibitem{dav}A. De Angelis and L. Vitale, CERN preprint CERN-OPEN 98-020, e-print
hep-ex/9808002.
\bibitem{delphi1}DELPHI Collaboration, P. Abreu {\sl at al}, {\it Phys. Lett.} {B} 
{\bf 286} {(1992)} {201}.
\bibitem{delphi2}DELPHI Collaboration, P. Abreu {\sl at al}, {\it Z. Phys.} {C} 
{\bf 63} {(1994)} {17}.
\end{thebibliography}
\end{document}